\documentclass[12pt,twoside]{article}
\usepackage{amsmath,amsfonts,amssymb}
\usepackage{latexsym}
\usepackage{dcolumn}
\usepackage{graphicx,epsfig}
\usepackage{amsthm}
\usepackage{hyperref}

\begin{document}

\begin{center}
{\Large \bf Black Hole Entropy with minimal length in Tunneling formalism}
\vglue 0.5cm
Barun Majumder\footnote{barunbasanta@iitgn.ac.in}
\vglue 0.6cm
{\small Indian Institute of Technology Gandhinagar \\ Ahmedabad, Gujarat 382424\\ India}
\end{center}
\vspace{.1cm}

\begin{abstract} 
Here we study the effects of the Generalized Uncertainty Principle in the tunneling formalism for Hawking radiation to evaluate the quantum-corrected Hawking temperature and entropy for a Schwarzschild black hole. We compare our results with the existing results given by other candidate theories of quantum gravity. In the entropy-area relation we found some new correction terms and in the leading order we found a term which varies as $\sim \sqrt{Area}$. We also get the well known logarithmic correction in the sub-leading order. We discuss the significance of this new quantum corrected leading order term.
\vspace{5mm}\newline Keywords: black hole entropy, tunneling , generalized uncertainty principle
\end{abstract}
\vspace{1cm}

One of the greatest achievement in theoretical physics in the last century is the realization that black holes are thermodynamic entities with well defined entropy and temperature \cite{w1,w3}. With simple ingredients Bekenstein first showed that the entropy of a black hole is proportional to the area of its horizon $S=\frac{A}{4l_p^2}$, $l_p$ being the Planck length. On the other hand with the field theory methods in curved background Hawking had shown that Schwarzschild black holes can emit thermal radiation with a temperature $T_H = \frac{\hbar}{8\pi M}$, where M being the mass of the black hole. In the recent years both string theory and loop quantum gravity have gained enormous amount of success in the statistical explanation of the black hole entropy \cite{w5}. Both the theories predicted quantum corrections to the entropy-area relation. Combining all the predictions from various theories of quantum gravity \cite{w7,w8,w9,x1,w10} we can have the following expansive form for the quantum corrected entropy-area relation:
\begin{equation}
\label{e2}
S= \frac{A}{4l_p^2} ~+~ c_0~\ln \bigg(\frac{A}{4l_p^2}\bigg) ~+~ \sum_{n=1}^\infty ~c_n \bigg(\frac{A}{4l_p^2}\bigg)^{-n} ~+~ \mathit{const.}~~,
\end{equation}
where the coefficients $c_n$ are model dependent parameters. Many researchers did an enormous amount of research in fixing the coefficient of the subleading logarithmic term $c_0$ () \cite{w7}. Recent rigorous calculations in the framework of loop quantum gravity predict $c_0$ to be $-1/2$ \cite{w10}.
\par
Quite recently there were some proposals to view Hawking radiation as a tunneling process in semi-classical quantum mechanics \cite{r6,r7}. Hawking radiation in a Schwarzschild background has been compared to Schwinger pair production, but there are important differences as particle pair production in a constant background electric field is in flat space-time. In the formalism developed in
\cite{r6,r7} the particles can follow the classically forbidden paths just after crossing the horizon. The tunneling amplitude depends
on the single particle classical action which becomes complex for the outgoing particle only. The calculation involves the imaginary part of the action which is related to the Boltzmann factor for the s-wave emission process across the horizon at the Hawking temperature. Using a semi-classical method based on complex path analysis one can show particle production in Schwarzschild like spacetimes with a horizon thereby recovering the Hawking radiation. Srinivasan and Padmanabhan \cite{r7} applied the Hamilton-Jacobi method for the computation of the imaginary part of the action. Simultaneously Parikh and Wilczek \cite{r6} used the method of radial null geodesic for the same purpose which was later used by authors in \cite{r8} for calculating the Hawking temperature for different spacetimes. Recent developments although semi-classical
include the same scheme for Dirac particles \cite{r9}. Authors in \cite{r10,rr1,rr2} made rigorous attempts for a detailed analysis. One can include the effects of back reaction and get the quantum corrections to the temperature and entropy \cite{r12}. An extension of the analysis
for non-commutative Schwarzschild spacetime was studied in \cite{r13}.
\par
Importance of the Generalized Uncertainty Principle (GUP) can be realized on the basis of simple \textit{gedanken} experiments without any reference of a particular theory of quantum gravity \cite{b2}. So we can think of the GUP as a model independent concept and relevant for the study of black hole entropy. Mead \cite{w11} first intuitively predicted that the Heisenberg's uncertainty principle could be affected by gravity. Later modified commutation relations between position and momenta commonly known as the GUP were shown to be present in different fundamental theories which claim to be a theory of quantum gravity. For example modified commutation relations between position and momenta were given by string theory, Doubly Special Relativity (DSR) with the prediction of a minimum measurable length \cite{b1,b2,b3}. Similar kind of commutation relation can also be found in the framework of Polymer Quantization in terms of polymer mass scale \cite{b4}. The GUP has been used by many authors for a heuristic analysis of the black hole entropy - see for example \cite{w13,w8,x2,w12,w13a}. For some important earlier developments see \cite{paddy1a} where a minimal length was introduced in a Lorentz invariant manner with some applications to thermal radiation. The authors in \cite{b5} proposed a GUP (ideally important for phenomenological purposes) which is consistent with DSR theory, string theory and black hole physics and which says
\begin{equation}
\left[x_i,x_j\right] = \left[p_i,p_j\right] = 0 ~~,
\end{equation}
\begin{equation}
\label{e4}
[x_i, p_j] = i \hbar \left[  \delta_{ij} -  l  \left( p \delta_{ij} + \frac{p_i p_j}{p} \right) + l^2  \left( p^2 \delta_{ij}  + 3 p_{i} p_{j} \right) \right]~~,
\end{equation}
\begin{align}
\label{e5}
 \delta x ~\delta p &\geq \frac{\hbar}{2} \left[ 1 - 2 l \langle p \rangle + 4 l^2 \langle p^2\rangle \right]  \nonumber \\
& \geq \frac{\hbar}{2} \left[ 1  +  \left(\frac{l}{\sqrt{\langle p^2 \rangle}} + 4 l^2  \right)  (\delta p)^2  +  4 l^2 \langle p \rangle^2 -  2 l \sqrt{\langle p^2 \rangle}~ \right],
\end{align}
where $ l=\frac{l_0 l_{p}}{\hbar} $ and $ l_{p} $ is the Plank length ($ \approx 10^{-35} m $). It is normally assumed that the dimensionless parameter $l_0$ is of the order unity which is not relevant for quantum gravity phenomenology. This assumption makes the $l$ dependent terms important at or near the Plank regime. But here we expect the existence of a new intermediate physical length scale of the order of $l \hbar = l_0 l_{p}$ and this unobserved length scale cannot exceed the electroweak length scale \cite{b5} which implies $l_0 \leq 10^{17}$. The above equations are approximately covariant under DSR transformations but not Lorentz covariant \cite{b3}. These equations also imply
\begin{equation}
\delta x \geq \left(\delta x \right)_{min} \approx l_0\,l_{p}
\end{equation}
and
\begin{equation}
\delta p \leq \left(\delta p \right)_{max} \approx \frac{M_{p}c}{l_0}
\end{equation}
where $ M_{p} $ is the Plank mass and $c$ is the velocity of light. With a lower bound for position fluctuations it is claimed that there is a minimum measurable distance and from an upper bound of momentum fluctuations we claim that momentum measurements cannot be arbitrarily imprecise. Recently, many authors have suggested \cite{dasprl,nat,bek} that the GUP implications can be measured directly in tabletop experiments which will definitely confirm the theoretical predictions of some models. If not everything then also we can get some experimental bound on the deformation parameter $l_0$. The effect of this proposed GUP is well studied recently for some well known physical systems in \cite{dasprl,b5,b6,w17}.\par
In this work we study the implications of the Generalized Uncertainty Principle in the tunneling formalism to evaluate the quantum-corrected Hawking temperature and entropy for a Schwarzschild black hole. We compare our results with the existing results in the literature as we found some new corrections to the temperature and the entropy-area relation. We conclude with some comments and discussion.\par
We start with the formalism developed in \cite{r6} (the method with radial null geodesic) which allows one to view Hawking radiation as quantum tunneling. For our purpose let us consider a general class of spherically symmetric static spacetime with line element
\begin{equation}
\label{r2.1}
ds^2 = - f(r) ~dt^2 + \frac{dr^2}{g(r)} + r^2d\Omega^2
\end{equation}
with the horizon located at $r=r_H$ such that $f(r)=g(r)=0$. In the Painleve coordinates \cite{r31} there is no singularity at the horizon but the metric do not remain static as in those coordinates (\ref{r2.1}) becomes \footnote{Although there are reasons to believe that Painleve coordinates are not good for this particular problem as they have two time contributions. The first time contribution can be seen from equation (1) of \cite{r6} where the relation between Schwarzschild time and Painleve time is given. After horizon crossing the argument of the log term becomes negative and gives an imaginary contribution. The emergence of second time contribution was shown in \cite{d2453} which is very crucial in getting the correct Hawking temperature preserving canonical invariance and unitarity.}
\begin{equation}
ds^2 = - f(r) ~dt^2 + 2~f(r)~\sqrt{\frac{1-g(r)}{f(r)~g(r)}}dt dr+dr^2 + r^2d\Omega^2 ~~.
\end{equation}
For the radial null geodesic we can find
\begin{equation}
\dot{r} = \sqrt{\frac{f(r)}{g(r)}}~(\pm 1 - \sqrt{1-g(r)})
\end{equation}
where $+(-)$ corresponds to outgoing (incoming) geodesics. $f(r)$ and $g(r)$ can be expanded about the horizon $r_H$ and this helps us to approximately write $\dot{r}$ as
\begin{equation}
\dot{r} \simeq \frac{1}{2} \sqrt{f'(r_H)~g'(r_H)}~(r-r_H) ~~.
\end{equation}
The imaginary part of the action can be written as \cite{r6}
\begin{equation}
\text{Im} ~S = \text{Im} \int_{r_{in}}^{r_{out}}~\int_{r_{H}}^{r_{0}}~ \frac{d H'}{\dot{r}}~dr~~.
\end{equation}
If we equate the tunneling probability ($\Gamma \sim e^{-\frac{2}{\hbar}~\text{Im}~S}$) with the Boltzmann factor ($e^{-\frac{\omega}{T}}$)
we can get the Hawking temperature as 
\begin{equation}
T_H = \frac{\omega \hbar}{2~\text{Im}~S} = \frac{\hbar}{4\pi}\sqrt{f'(r_H)g'(r_H)}
\end{equation}
which for a Schwarzschild black hole is $\frac{\hbar}{8\pi M}$. There is a `factor of 2' ambiguity as discussed in \cite{r14} as $\oint p ~dr$ is invariant under canonical transformations but $\int p~dr$ is not. Now we will apply the Hamilton-Jacobi method as developed in \cite{r7} for the calculation of the imaginary part of the action and thereby calculate the Hawking temperature and entropy for a Schwarzschild black hole. The method has an advantage over the radial null geodesic method as this is valid for massive particles. For simplicity we consider a massless particle in the spacetime described by eqn. (\ref{r2.1}). The Klein-Gordon equation is given by
\begin{equation}
-\frac{\hbar^2}{\sqrt{-g}}~ \partial_{\mu} \left(g^{\mu\nu}\sqrt{-g}~\partial_{\nu}\right)\phi = 0 ~~.
\end{equation}
We can use the ansatz 
\begin{equation}
\label{r3.3}
\phi (r,t) = e^{-\frac{i}{\hbar}~S(r,t)}
\end{equation}
for the semi-classical wave function of the Klein-Gordon equation. To incorporate the quantum corrections in powers of $\hbar$ we can expand
$S(r,t)$ as
\begin{equation}
\label{r3.5}
S(r,t) = S_0 (r,t) + \sum_i \hbar^i S_i (r,t)~~,
\end{equation}
where $S_0$ is the semi-classical value of the action. Using dimensional argument we write the form of eqn. (\ref{r3.5}) as 
\begin{equation}
\label{new18}
S(r,t) = \left[ 1+ \sum_i \beta_i \frac{\hbar^i}{M^{2i}} \right]~S_0(r,t)
\end{equation}
where M is the mass of the black hole and considering the choice of unit where $G=c=k_B=1$ with $\beta_i$'s being the dimensionless constant parameters. While writing (\ref{new18}) we made an assumption that all higher order corrections are proportional to $S_0$. We considered the spacetime of (\ref{r2.1}) as static and has timelike killing vectors so we would like to go for a solution \cite{r7}
\begin{equation}
\label{om1}
S_0(r,t) = \omega t + \tilde{S_0}(r)
\end{equation}
where $\omega$ is the energy of the particle (see Appendix). Now it is interesting to note that we are studying a process that occurs at the horizon and so this energy of the particle should receive some corrections due to the strength of gravity. Though the argument is heuristic but still we have some reason to believe this. Recently in \cite{li1} it is shown that the effect of the generalized uncertainty principle becomes more and more important as one approaches the event horizon. Now we see that eqns. (\ref{e4}) and (\ref{e5}) represents modified Heisenberg algebra. The interesting part of these two relationships is the term which is linear in $l(=l_0l_p/\hbar)$ with $p$. This term is actually important for phenomenological purposes and for our purpose we will consider the generalized Heisenberg algebra (Generalized Heisenberg principle) with a small change in notation where $x$ and $p$ obeys the relation ($\alpha >0$)
\begin{equation}
\label{e8}
\delta x ~\delta p \geq \hbar~ \bigg[~1 - \frac{\alpha l_p}{\hbar} ~\delta p + \frac{\alpha^2 l_p^2}{\hbar^2} ~(\delta p)^2~\bigg] ~~.   
\end{equation}  
While writing equation (\ref{e8}) we have made an approximation that $(\delta p)\approx \sqrt{\langle p^2 \rangle}$. Eventually this
would mean $\langle p \rangle \approx 0$. For the study of Schwarzschild black hole which is spherically symmetric this seems to be a valid approximation \footnote{In many problems of quantum mechanics we usually find $\langle p \rangle =\langle x \rangle =0~$ (for example for the ground state of harmonic oscillator).}. We can see that if $\alpha =2l_0$ this is the same relation as that of (\ref{e5}). Here $\delta p$ and $\delta x$ are the momentum and position uncertainty for a quantum particle and $\alpha$ is a dimensionless positive parameter (also known as deformation parameter in the literature of non-commutative geometry).
As $l_p=\sqrt{\frac{\hbar G}{c^3}}$, where $G$ is the Newtonian coupling constant, we can comment that the extra terms in the uncertainty relation is due to incorporation of gravity. We now re express the GUP of (\ref{e8}) in the following form
\begin{equation}
\label{e9}
\delta p \geq \frac{~\hbar~(\delta x + \alpha l_p) ~-~ \hbar \sqrt{(\delta x + \alpha l_p)^2 - 4\alpha^2 l_p^2}~}{2\alpha^2 l_p^2}~~,
\end{equation}
where we have chosen a negative sign by taking the classical limit. Normally $l_p$ is viewed as an ultraviolet cut-off on spacetime geometry (e.g.,\cite{w14}), so it is quite justified and relevant that we can consider the dimensionless ratio $\frac{l_p}{\delta x}$ relatively small as compared to unity. Now we Taylor expand equation (\ref{e9}) and express the same equation with some simple manipulation as
\begin{equation} 
\delta p \geq \frac{1}{\delta x} \bigg[1 - \frac{\alpha }{2(\delta x)} + \frac{\alpha^2 }{2(\delta x)^2} - \frac{\alpha^3 }{2(\delta x)^3}
+ \frac{9}{16}\frac{\alpha^4 }{(\delta x)^4} - \ldots \bigg] ~~.
\end{equation} 
So far we have worked in units where $G=c=k_B=1$ and from now on we also choose $\hbar=1$, so we have $l_p=1$. The Heisenberg uncertainty principle ($\delta p \delta x \geq 1$) can be translated to the lower bound $\omega\delta x \geq 1$ with the arguments used in \cite{w15,x1}, where $\omega$ is the energy of a quantum particle. If we imply our GUP, we can rebuild the lower bound as
\begin{equation}
\omega_G \geq \omega \bigg[1 - \frac{\alpha }{2(\delta x)} + \frac{\alpha^2 }{2(\delta x)^2} - \frac{\alpha^3 }{2(\delta x)^3} + \frac{9}{16}\frac{\alpha^4 }{(\delta x)^4} - \ldots \bigg] ~~,
\end{equation}
where $\omega_G$ is the GUP corrected energy. So now with the mentioned arguments we re-write eqn. (\ref{om1}) as
\begin{equation}
\label{r3.9a}
S_0(r,t) = \omega_G ~t + \tilde{S_0}(r) ~~.
\end{equation}
With eqn. (\ref{new18}) and  (\ref{r3.9a}) we can write
\begin{equation}
\tilde{S_0} (r) = \pm \omega_G ~ \int_0^r ~\frac{dr}{\sqrt{f(r)~g(r)}} ~~.
\end{equation}
+ (-) denotes the incoming (outgoing) of the particle. So now
\begin{equation}
\label{r3.11}
S(r,t) = \left( 1 + \sum_i \beta_i \frac{1}{M^{2i}} \right)~\left(\omega_G ~ t \pm \omega_G ~\int_0^r ~\frac{dr}{\sqrt{f(r)~g(r)}}
\right) ~~.
\end{equation}
Now we can write the solutions of the Klein-Gordon equation for the incoming and outgoing part with eqns. (\ref{r3.3}) and (\ref{r3.11}) as
\begin{equation}
\phi_{in} = \text{exp} ~\left[-i\left( 1 + \sum_i \beta_i \frac{1}{M^{2i}} \right)
\left(\omega_G ~t + \omega_G ~\int_0^r ~\frac{dr}{\sqrt{f(r)~g(r)}}\right) \right] 
\end{equation}
and
\begin{equation}
\phi_{out} = \text{exp} ~\left[-i\left( 1 + \sum_i \beta_i \frac{1}{M^{2i}} \right)
\left(\omega_G ~t - \omega_G ~\int_0^r ~\frac{dr}{\sqrt{f(r)~g(r)}}\right) \right] ~~.
\end{equation}
The incoming and outgoing probabilities of the particle can be calculated from the solutions of the Klein-Gordon equation and are given by
\begin{equation}
\label{r3.14}
P_{in} = \vert \phi_{in} \vert ^2 = \text{exp} ~\left[2\left( 1 + \sum_i \beta_i \frac{1}{M^{2i}} \right)
\left(\omega_G ~\text{Im}~t + \omega_G ~\text{Im}~\int_0^r ~\frac{dr}{\sqrt{f(r)~g(r)}}\right) \right] 
\end{equation}
and
\begin{equation}
\label{r3.15}
P_{out} = \vert \phi_{out} \vert ^2 = \text{exp} ~\left[2\left( 1 + \sum_i \beta_i \frac{1}{M^{2i}} \right)
\left(\omega_G ~\text{Im}~t - \omega_G ~\text{Im}~\int_0^r ~\frac{dr}{\sqrt{f(r)~g(r)}}\right) \right] ~~.
\end{equation}
For tunneling of a particle through the horizon the temporal coordinates suffers a rotation in the complex plane and hence eqns. (\ref{r3.14}) and (\ref{r3.15}) shows the corresponding contribution from the imaginary part. It is very important that we take into account the rotation for the temporal part of the action which leads to an imaginary contribution. The total imaginary contribution finally yields the standard Hawking temperature. Following \cite{d2453,dplb} we get 
\begin{equation}
\text{Im}~t = - ~\text{Im}~\int_0^r ~\frac{dr}{\sqrt{f(r)~g(r)}}
\end{equation}
and
\begin{equation}
P_{out} = \text{exp} \left[ - 4~\omega_G~\left( 1 + \sum_i \beta_i \frac{1}{M^{2i}} \right)~\text{Im}~
\int_0^r ~\frac{dr}{\sqrt{f(r)~g(r)}} \right]~~.
\end{equation}
Following the argument in \cite{r7} we can give a thermal interpretation to the result by comparing with the Boltzmann
factor ($e^{-\frac{\omega}{T}}$) and eventually obtain the quantum corrected temperature of the black hole
\begin{equation}
\label{thr}
T = \frac{T_H}{\left(1 - \frac{\alpha }{2(\delta x)} + \frac{\alpha^2 }{2(\delta x)^2} - \frac{\alpha^3 }{2(\delta x)^3}
 + \ldots \right) \left( 1 + \sum_i \beta_i \frac{1}{M^{2i}} \right)} ~~,
\end{equation}
where $T_H = \frac{1}{4}~\left(\text{Im}~\int_0^r ~\frac{dr}{\sqrt{f(r)~g(r)}}\right)^{-1}$ is the standard Hawking temperature with specific f(r) and g(r). For a Schwarzschild black hole $f(r)=g(r)=(1-\frac{2M}{r})$ and $r_H=2M$ and hence the standard Hawking temperature
is found to be $T_H = \frac{1}{8\pi M}$. Here we will choose $\delta x \sim 2r_H = 4M $ (a brief argument can be found in \cite{w16,x2}). If we put this in eqn. (\ref{thr}) and Taylor expand the expression we get the quantum corrected temperature for the black hole as 
\begin{equation}
T = T_H \left[1 + \frac{\alpha }{8M} - \left(\frac{\alpha^2}{32} + \beta_1 \right) \frac{1}{M^2} + \ldots \ldots ~~~~~~~~
{\cal O}~\left(\frac{1}{M^3} ~\text{or higher}\right) \right] ~~.
\end{equation}
Already we have some existing results in the literature. With one loop back reaction effect this modification of the Hawking temperature was obtained in \cite{r12}. Similar results can also be obtained with techniques from conformal field theory \cite{r22}. There the prefactor of the quadratic term in $\frac{1}{M}$ is related to the trace anomaly ($\alpha'$) which depends on the number of fields with specific spin.
But in addition here we have obtained a new correction term ($\frac{\alpha }{8M}$) which is entirely different from the existing results contributing positively to the standard Hawking temperature.\par
Using the law of black hole thermodynamics we get the entropy of a Schwarzschild black hole as 
\begin{equation}
\label{bht}
S_{bh} = \int \frac{dM}{T} ~~.
\end{equation}
Putting (\ref{bht}) in (\ref{thr}) we get the quantum corrected entropy as
\begin{align}
\label{fin}
S_{bh} &= \frac{A}{4} - \frac{\sqrt{\pi}\alpha}{2}~ \sqrt{\frac{A}{4}} + 4\pi \left(\beta_1 + \frac{\alpha^2}{32}\right) \ln \frac{A}{4} + 2 \pi^{3/2} \alpha \beta_1 \frac{1}{\sqrt{\frac{A}{4}}} - \frac{\pi^2 \alpha^2 \beta_1}{2} \frac{1}{\frac{A}{4}} 
+  \ldots\ldots ~~,
\end{align}
where $A=4 \pi r_H^2=16 \pi M^2$ is the area of the event horizon. $\frac{A}{4}$ is the standard Bekenstein-Hawking entropy. We rewrite
eqn. (\ref{fin}) in the form of an expansion as
\begin{align}
S  \simeq & \frac{A}{4} - \frac{\pi^{1/2} \alpha}{2}\sqrt{\frac{A}{4}} + 4\pi \left(\beta_1 + \frac{\alpha^2}{32}\right) \ln \frac{A}{4} \nonumber \\
& + \sum_{m=\frac{1}{2},\frac{3}{2},\ldots}^\infty d_m  \left(\frac{A}{4}\right)^{-m} - 
\sum_{n=1,2,\ldots}^\infty c_n  \left(\frac{A}{4}\right)^{-n} + \mathit{const.} ~~.
\end{align}
Here $m$ denotes positive half-integers and $n$ positive integers. If we compare this equation with (\ref{e2}) we can see that there are extra terms in this equation. One of the leading contribution to the entropy is from the new second term $\sim \sqrt{\mathit{Area}}$. Also we have other new correction terms proportional to $(\mathit{Area})^{-m}$. In \cite{w13} black hole thermodynamics was first studied with modified dispersion relations and the generalized uncertainty principle yielding the same result. This was also pointed out in \cite{w17} and later in \cite{mybh} for the same situation.\par
So in this work we study the effects of the Generalized Uncertainty Principle in black hole tunneling formalism as recently developed. We applied the Hamilton-Jacobi method for the calculation of the imaginary part of the action and the GUP is introduced through the correction to the energy of a particle due to gravity in the immediate vicinity of the horizon. Later we calculated the quantum corrected temperature for a Schwarzschild black hole and found some new correction terms as compared to the existing results in the literature. The effect of these new corrections remained in the expression of the Bekenstein-hawking entropy and the leading order correction being $\sim \sqrt{\text{A}}$, where A is the area of the event horizon. Our leading order correction is different and hence do not agree with the entropy bound as given by the local quantum field theory or the holographic principle \cite{rama1,rama3}. Although it can be shown that the holographic principle (with regard to cosmology) will remain valid  as long as our universe (if flat or open) is \textit{non-planckian}.  The philosophy of interpretation of the new leading order $\sqrt{\text{A}}$- type correction in black-hole entropy also seems to be awaiting a new direction.

\subsubsection*{\underline{Appendix}}

Let us consider a static line element in $(1+1)$ dimension
\begin{equation}\label{a1}\tag{a1}
ds^2 = f(r)dt^2 - \frac{1}{f(r)}dr^2
\end{equation}
where $f(r)$ is an arbitrary function of $r$. We also consider that at $r=r_0$ there is a horizon and $f(r)\vert_{r=r_0}=0$. If we take a massive minimally coupled scalar field $\phi$ in this background then the field $\phi$ satisfies the Klein-Gordon equation
\begin{equation}\tag{a2}
\left[\Box + \frac{m^2}{\hbar^2}\right]\phi = 0 ~~.
\end{equation}
Evaluating $\Box$ operator in the background \ref{a1} we can write
\begin{equation}\tag{a3}
\frac{1}{f(r)}\frac{\partial^2 \phi}{\partial t^2} - \frac{\partial}{\partial r} \left\{f(r)\frac{\partial \phi}{\partial r}\right\}=-\frac{m^2}{\hbar^2}\phi ~~.
\end{equation}
We can rewrite this equation as
\begin{equation}\tag{a4}
\left[\frac{1}{f(r)}\left(\frac{\partial S}{\partial t}\right)^2 - f(r)\left(\frac{\partial S}{\partial r}\right)^2 -m^2 \right] - i\hbar
\left[\frac{1}{f(r)}\frac{\partial^2 S}{\partial t^2} - f(r) \frac{\partial^2 S}{\partial r^2} - \frac{df(r)}{dr}\frac{\partial S}{\partial r}\right] = 0
\end{equation}
where we considered the ansatz $\phi (r,t) = e^{\frac{i}{\hbar} S(r,t)}$ for the semi-classical wave function of the Klein-Gordon equation. If we expand $S(r,t)$ in powers of ($\frac{\hbar}{i}$) then $S_0(r,t)$, the leading order in the expansion, satisfies the Hamilton-Jacobi equation
\begin{equation}\tag{a5}
\frac{1}{f(r)} \left(\frac{\partial S_0}{\partial t}\right)^2 - f(r)\left(\frac{\partial S_0}{\partial r}\right)^2 - m^2 = 0 ~~.
\end{equation}
The solution of this equation is
\begin{equation}\label{a6}\tag{a6}
S_0(r,t) = -Et \pm \int \frac{dr}{f(r)}\sqrt{E^2 - m^2 f(r)} ~~.
\end{equation}
In the massless case the solution is
\begin{equation}\tag{a7}
S_0(r,t)\vert_{m=0} = f_1 (t-r_t) + f_2 (t+r_t)
\end{equation}
where $r_t=\int \frac{dr}{f(r)}$ is the {\it tortoise} coordinate and $f_1$ and $f_2$ are arbitrary functions. The choice of $f_{1,2} = -Et\pm Er_t$
gives eqn. \ref{a6} in the massless case. Here $E$ is identified with energy. So we can see that the semiclassical ansatz is exact for the massless scalar field but one can also show that the final results remain same for the massive scalar field \cite{r7}.

\section*{Acknowledgements}
The author was partly supported by the Excellence-in-Research Fellowship of IIT Gandhinagar. The author would like to thank Prof. Douglas Singleton for enlightening comments and helpful suggestions which helped immensely in writing a major part of the manuscript. The author would also like to thank an anonymous referee for suggestions.


\end{document}